\documentclass[aps,prl,preprint,groupedaddress]{revtex4-1}
\usepackage{color}
\draft 
\usepackage{graphicx}
\usepackage{caption}
\usepackage{float}
\usepackage{textcomp}

\begin{document}


\title{Correlation between structural defects and optical properties of $\mathrm{Cu_2O}$ nanowires grown by thermal oxidation}


\author{Qilin Gu}
\email[]{gump423@gmail.com}
\homepage[]{https://sites.google.com/site/qlgu423/}
\affiliation{Department of Electrical and Computer Engineering, The Ohio State University, Columbus, OH 43210, USA}
\author{B. Wang}
\email[]{bwang018@ansteel.com.cn}
\affiliation{Anshan Iron \& Steel group Co. Ltd. Technical Center, Anshan 114021, Liaoning, China}


\date{December 3, 2010}

\begin{abstract}
Cuprous oxide ($\mathrm{Cu_2O}$) nanowires were grown by a simple, catalyst-free thermal oxidation method starting from a copper foil substrate. Wire growth was performed at different conditions by varying oxygen partial pressure and heating temperature, and produced structural defects such as stacking faults and twin structures of grown nanowires were characterized by transmission electron microscopy. A moderate oxygen partial pressure $\scriptsize{\sim}$5\% with a relatively high temperature at 500 \textcelsius{} is found to be an optimal condition for growing high-quality $\mathrm{Cu_2O}$ nanowires, while either lowering temperature or enhancing $\mathrm{O_2}$ partial pressure can yield higher density of planar defects. In order to further investigate optical properties of grown nanowires, photoluminescence measurements were also performed and correlated with structural characterization results, indicating that increase of structural defects such as stacking faults and twins tended to enhance broad defect-related emissions and suppress the exciton emissions near $\mathrm{Cu_2O}$ band-edge. 
\end{abstract}


\maketitle

As the first demonstrated semiconductor material, copper (I) oxide ($\mathrm{Cu_2O}$) has been regarded as one of the most promising candidates for various optoelectronic applications including rectifying diodes, photovoltaics, non-volatile memories and spintronics.\textcolor{blue}{$^{1-2}$} Despite the decline of the attention on $\mathrm{Cu_2O}$ due to the emergence of silicon and germanium and their dominations in modern electronics, $\mathrm{Cu_2O}$ is always of great research interests due to the ease of its synthesis process, especially in the field of low-cost inorganic solar cells.\textcolor{blue}{$^{3}$} The particular advantages of $\mathrm{Cu_2O}$ as a photovoltaic material lie in its good absorption coefficient for the light above the bandgap, its reasonably good majority carrier mobility and minority carrier diffusion length, and the abundant amount of available composing elements with non-toxic natures.\textcolor{blue}{$^{4}$}

Rational schemes for $\mathrm{Cu_2O}$ synthesis and subsequent material characterizations have become increasingly crucial in order to realize the corresponding desired applications. Although most early studies have focused on bulk materials, there is expanding interest in producing one-dimensional $\mathrm{Cu_2O}$-based nanowires (NWs) and nanorods (NRs) structures which are of high aspect ratios owing to their remarkable physical properties such as expected quantum effect and potentially high crystalline quality. 1-D $\mathrm{Cu_2O}$ nanostructures have been synthesized by using a variety of methods including redox reactions in $\mathrm{Cu^{2+}}$-based solutions,\textcolor{blue}{$^{5-7}$} electrochemical deposition through alumina templates,\textcolor{blue}{$^{8-10}$} most of which were solution-based techniques. As an alternative feasible route, direct thermal oxidation method for producing oxide nanowires has become attractive for its simplicity and been successfully adopted in synthesis of $\alpha$-$\mathrm{Fe_2O_3}$,\textcolor{blue}{$^{11-12}$} ZnO,\textcolor{blue}{$^{13-14}$} $\beta$-$\mathrm{Ga_2O_3}$\textcolor{blue}{$^{15-16}$} and even CuO/$\mathrm{Cu_2O}$\textcolor{blue}{$^{17-18}$} nanowires. In this letter, a similar oxidation method for growing $\mathrm{Cu_2O}$ NWs through heating metal substrate directly under controlled ambient is demonstrated. The structural defects in synthesized nanowires under different conditions have been characterized by electron microscopy, which have also been correlated with their corresponding optical properties obtained from photoluminescence studies.

A 2"-diameter horizontal quartz tube furnace (Barnstead/Thermolyne-F21135) with ceramic fiber insulation for rapid heat-up and cool-down processes was utilized as the major part of the thermal oxidation synthesis system. An additional gas regulator was attached to the furnace to precisely control the flow rates for desired gas conditions. Source material copper foil (99.99\% purity, 0.1 mm thick, Tongchuang Copper Ltd., China) substrate was first dipped into HCl solution for 10 sec, followed by cleaning and degreasing in ultrasonic bath for 10 minutes and subsequent repeated rinsing in deionized (DI) water. After being dried under nitrogen ($\mathrm{N_2}$) flow, the substrate was then placed in the tube furnace on an alumina ceramic boat with 100 sccm Ar flowing in the tube furnace to evacuate residual air prior to nanowire growth. Once 30-minute Ar flushing ended, the gas flow rate was adjusted to proposed values, typically a relatively slow flow rate of 25 sccm for $\mathrm{O_2}$ and Ar mixture, for nanowire growth. Upon finishing the oxidation process, 10\% $\mathrm{H_2}$ was then introduced to replace $\mathrm{O_2}$ for reduction of obtained CuO phase at a lower temperature of 200 \textcelsius{} for 30 minutes. Multiple synthesis tests have been carried out by varying parameters including furnace temperature, growth time and $\mathrm{O_2}$ partial pressure in mixed gas to study their impacts on properties of synthesized nanowires. To investigate the morphology and structural properties of $\mathrm{Cu_2O}$ NWs, a field-emission scanning electron microscope (FESEM) combined with EDX system (SSX-550, Shimadzu) was used to capture SEM images and extract chemical compositional information. TEM and high-resolution TEM (HRTEM) images were acquired with a JEOL 2000FX microscope operating at 200 \emph{kV} and a JEOL JEM 4000FX operating at 350 \emph{kV} respectively. X-ray diffraction (XRD) patterns were recorded with a Shimadzu XRD-6000 diffractometer with Cu K$\alpha$ radiation ($\lambda$=1.5406 \AA). Furthermore, optical emission properties of synthesized $\mathrm{Cu_2O}$ NWs were also studied by photoluminescence (PL) measurements in the wavelength range of 550-1050 nm at the liquid nitrogen temperature (77K). The 488 nm line of an $\mathrm{Ar^+}$ laser was used to excite the luminescence in the nanowire samples and the signal was collected by a detector consisting of a Zolix grating spectrometer with a Hamamatsu R5108 photo-multiplier tube (PMT).

A plan-view SEM image displaying typical surface morphology in $\mathrm{Cu_2O}$ nanowires produced at 320 oC with 5\% $\mathrm{O_2}$ flow is shown in Figure 1(a). A bundle of uniform wire-shape nanostructures with diameters of 30-40 nm and lengths of a few microns can be clearly observed. Morphological dependence on oxidation time indicates that growth time of 5-7 hours yields dense nanowires bundle and optimal surface morphology with excellent uniformity of wire diameters, so the thermal synthesis time was fixed to be 5 hours for all the subsequent studies. The EDX result at the inset of Fig. 1(a) confirms that the product is only composed of two elements Cu and O with atomic ratio of roughly 2:1. A typical TEM image of a single $\mathrm{Cu_2O}$ nanowire in Fig. 1(b) shows a relatively smooth surface of the grown structure with an almost identical diameter along the wire length. Selected area electron diffraction (SAED) pattern is shown in the inset of Fig. 1(b). The exactly same SAED patterns were obtained from different parts of the nanowire, indicating their single crystallinity. Electron diffraction has been performed on tens of wires, and the corresponding patterns suggest that the growth of nanowires was along either $<$111$>$, $<$110$>$ or $<$001$>$ direction.

The most common planar defects in 1-D inorganic nanowires, such as twins, stacking faults, and even inversion domain walls, are not only essential for the growth of the nanostructures, but also strongly affect their optical and electrical properties.\textcolor{blue}{$^{19}$} In this study, major planar defects have been studied by electron microscopy methods. The right column of Figure 2 shows a series of HRTEM images of selected regions that exhibit representative structural features in the single nanowire shown in the left column. The upper image represents a nearly defect-free spot where stacking faults or twins don't exist. One can observe that atoms in this region are perfectly arranged and ordered stacked without any errors. The lattice fringes, according to HRTEM images, were separated by 0.25 nm, in good agreement with the spacing between (111) planes. In contrast, a common twin structure that appears in produced $\mathrm{Cu_2O}$ nanowires is shown in the middle image, where a mirror-like highly symmetric interface was created between two ordered atomic domains. The rotation angle between twinning segments, as marked in the figure, is 141o, matching twice the inter-planar angle (70.5o) formed between two {111} planes for cubic structures.\textcolor{blue}{$^{20}$} This type of twin structures with angle of 141o, known as the first order twin, has been widely observed in various nanowires as they are more energetically favorable and consequently more stable then higher order twins.\textcolor{blue}{$^{21}$} The lower image in Figure 2 shows a typical structure of intrinsic stacking faults perpendicular to the $<$111$>$ growth direction, as indicated by the white arrow. In this spot, an interrupting discontinuity in the normal stacking sequence is present and a slightly shift of crystallographic planes with respect to each other can be observed due to the inclusion of this stacking fault plane into the regular cubic structure of $\mathrm{Cu_2O}$ nanowire.

To further explore the impacts of heating temperature and $\mathrm{O_2}$ content in mixing gas on planar defects in synthesized $\mathrm{Cu_2O}$ nanowires, HRTEM images taken from nanowires oxidized under three representative growth conditions are shown in Fig. 3(a)-(c). For the case for relatively low synthesis temperature of 320 \textcelsius{} and low $\mathrm{O_2}$ partial pressure of 5\% (Fig. 3(a)), moderate density of planar defects can be observed in a representative nanowire. When oxidation temperature increases to 500 \textcelsius{} and $\mathrm{O_2}$ content is kept at 5\% level, planar defects can hardly be seen along a single nanowire except an occasional one shown in Figure 3(b), indicating that the average density of stacking faults and twins drastically drops to a very low level. Further increase of temperature to 640 \textcelsius{} and meanwhile enhancing $\mathrm{O_2}$ partial pressure to 20\%, however, resulted in an exceedingly large quantity of planar defects. As a result, consequently formed high defect density network can be seen in Figure 3(c). Although it is believed that the observed variation of planar defect density is related to the impact of thermodynamic conditions on strain distributions during wire formation processes,\textcolor{blue}{$^{22}$} the corresponding detailed mechanism still needs further studies. Strong and sharp XRD peaks and the absence of Cu and CuO traces for all three samples shown in Fig. 3(d) suggest that the grown $\mathrm{Cu_2O}$ NWs are highly crystalline and purely in $\mathrm{Cu_2O}$ phase. The lower crystal quality of sample produced at 640 \textcelsius{} with 20\% $\mathrm{O_2}$ indicated by its broader diffraction peaks corresponds to the highest density of planar defects revealed by HRTEM results. Differences in relatively intensity ratio between individual diffraction peaks imply the anisotropic growth of these planes during the oxidation processes. 

Low-temperature (LT) photoluminescence results for samples synthesized under three growth conditions are individually presented in Figure 4(b) with their corresponding TEM images showing defect structures (Fig. 4(c)). The primary PL spectral characteristics for all three nanowire samples include two emission peaks centering at $\scriptsize{\sim}$590 nm and $\scriptsize{\sim}$750 nm respectively. The emission at $\scriptsize{\sim}$590 nm (2.1 eV) matches $\mathrm{Cu_2O}$ bandgap and is thereby attributed to near band-edge transition from free or bound exciton recombination.\textcolor{blue}{$^{23}$} On the other hand, a broad emission peak at red-near infrared (\emph{IR}) region, which is considered to be an overlapping luminescent effect from multiple defect-related radiative recombination centers and broadened by acoustic phonon bath dissipation, can also be observed in each $\mathrm{Cu_2O}$ NWs samples.\textcolor{blue}{$^{24}$} By normalizing all three PL spectra to the 488 nm line intensity (not shown here), the direct comparison (Fig. 4(a)) shows that the sample with extremely high density of defects exhibits significantly strong defect-related emission PL peak at near \emph{IR} region with an almost invisible exciton emission near $\mathrm{Cu_2O}$ band-edge, and that the optimal wires with the lowest defect density show strong exciton PL emission and greatly suppressed near-\emph{IR} emission. Note that the evident red-shift of the defect emission peak from nanowire grown at 640 \textcelsius{} with 20\% $\mathrm{O_2}$ (green dash line) is probably a consequence of oxygen vacancy decrease due to higher $\mathrm{O_2}$ content during the oxidation. 

In conclusion, $\mathrm{Cu_2O}$ nanowires have been successfully synthesized by a simple thermal oxidation method through heating copper foil under $\mathrm{O_2}$/Ar mixing gas ambient with subsequent $\mathrm{H_2}$ reduction process. Planar defects including stacking faults and twins are observed in grown nanowires under a variety of temperatures and $\mathrm{O_2}$ contents. Notably, heating substrate at 500 \textcelsius{} with a relatively low $\mathrm{O_2}$ partial pressure of 5\% is found to be an optimal oxidation condition under which very low density of planar defects can be achieved. Decreasing synthetic temperature generates moderately higher defect density and increasing both temperature and $\mathrm{O_2}$ content yields tremendously larger amount of twins and stacking faults. Optical properties from PL measurements correspond well to the structural defect features of nanowire samples for individual synthetic conditions. The established correlation suggests that structural defects can lead to accordingly considerable impacts on optical properties of $\mathrm{Cu_2O}$ nanowires grown by thermal oxidation methods.

\begin{acknowledgments}
The authors wish to thank Dr. P. L. Cai at Center for Inorganic Materials Engineering, Liaoning University of Science and Technology, for his help on TEM and SEM experiments
\end{acknowledgments}

\vfill
\newpage
\noindent
\underline{\textbf{Reference}}\\
$^1$E. H. Kennard and E. O. Dieterich, Phys. Rev. \textbf{9}, 58 (1917).\\
$^2$L. O. Grondahl and P. H. Geiger, J. Am. Inst. Electr. Eng. \textbf{46}, 215 (1927).\\
$^3$A. E. Rakhshani, Solid-State Electron. \textbf{29}, 7 (1986).\\
$^4$C. Wadia, A. P. Alivisatos and D. M. Kammen, Environ. Sci. Technol. \textbf{43}, 2072 (2009).\\
$^5$W.Z. Wang, G.H. Wang, X.S. Wang, Y.K. Liu and C.L. Zheng, Adv. Mater. \textbf{14}, 67 (2002).\\
$^6$Y. Tan, X. Xue, Q. Peng, H. Zhao, T. Wang and Y. Li, Nano Lett. \textbf{7}, 3723 (2007).\\
$^7$X. Liu, R. Hu, S. Xiong, Y. Liu, L. Chai, K. Bao and Y. Qian, Mater. Chem. Phys. \textbf{114}, 213 (2009).\\
$^8$J. Oh, Y. Tak and J. Lee, Electrochem. Solid-State Lett. \textbf{7}, C27 (2004).\\
$^9$X. M. Liu and Y. C. Zhou, Appl. Phys. A \textbf{81}, 68 (2005).\\
$^{10}$E. Ko, J. Choi, K. Okamoto, Y. Tak and J. Lee, ChemPhysChem \textbf{7}, 1505 (2006).\\
$^{11}$R. L. Tallman and E. A. Gulbransen, J. Electrochem. Soc. \textbf{114}, 1227 (1967).\\
$^{12}$P. Hiralal, H. E. Unalan1, K. G. U. Wijayantha, A. Kursumovic, D. Jefferson, J. L. MacManus Driscoll and G. A. J. Amaratunga, Nanotechnology \textbf{19}, 455608 (2008).\\
$^{13}$S. Ren, Y. F. Bai, J. Chen, S.Z. Deng, N.S. Xu, and S. Yang, Mater. Lett. \textbf{61}, 666 (2007).\\
$^{14}$S. Rackauskas, A. G. Nasibulin, H. Jiang, Y. Tian, V. I. Kleshch, J. Sainio, S. N. Bokova, A. N. Obraztsov, and E. I. Kauppinen, Nanotechnology \textbf{20}, 165603 (2009).\\
$^{15}$Z. R. Dai, Z. W. Pan and Z. L. Wang, J. Phys. Chem. B \textbf{106}, 902 (2002).\\
$^{16}$B. C. Kim, K. T. Sun, K. S. Park, K. J. Im, T. Noh, M. Y. Sung, S. Kim, S. Nahm, Y. N. Choi and S.S. Park Appl. Phys. Lett. \textbf{80}, 479 (2002).\\
$^{17}$C. C. Jiang, T. Herricks and Y. Xia, Nano Lett. \textbf{2}, 1332 (2002).\\
$^{18}$L. Liao, B. Yan, Y. F. Hao, G. Z. Xing, J. P. Liu, B. C. Zhao, Z. X. Shen, T. Wu, L. Wang, J. T. L. Thong, C. M. Li, W. Huang and T. Yu, Appl. Phys. Lett. \textbf{94}, 113106 (2009).\\
$^{19}$Y. Ding and Z. L. Wang, Micron \textbf{40}, 335 (2009).\\
$^{20}$D. Shechtman, A. Feldman, M.D. Vaudin, J.L. Hutchison, Appl.Phys.Lett. \textbf{62}, 487(1993).\\
$^{21}$T. L. Daulton, T. J. Bernatowicz, R. S. Lewis, S. Messenger, F. J. Stadermann and S. Amari. Geochimica et Cosmochimica Acta \textbf{67}, 4743 (2003).\\
$^{22}$D.-H. Wang, D. Xu, Q. Wang, Y.-J. Hao, G.-Q. Jin, X.-Y. Guo and K. N. Tu. Nanotechnology \textbf{19}, 215602 (2008).\\
$^{23}$B. Prevot, C. Carabatos and M. Sieskind, Phys. Status Solidi A \textbf{10}, 455 (1972).\\
$^{24}$S. V. Gastev, A. A. Kaplyanskii and N. S. Sokolov Solid State Commun. \textbf{42}, 389 (1982).\\

\begin{figure}[H]
\centering
\includegraphics[width=1\textwidth]{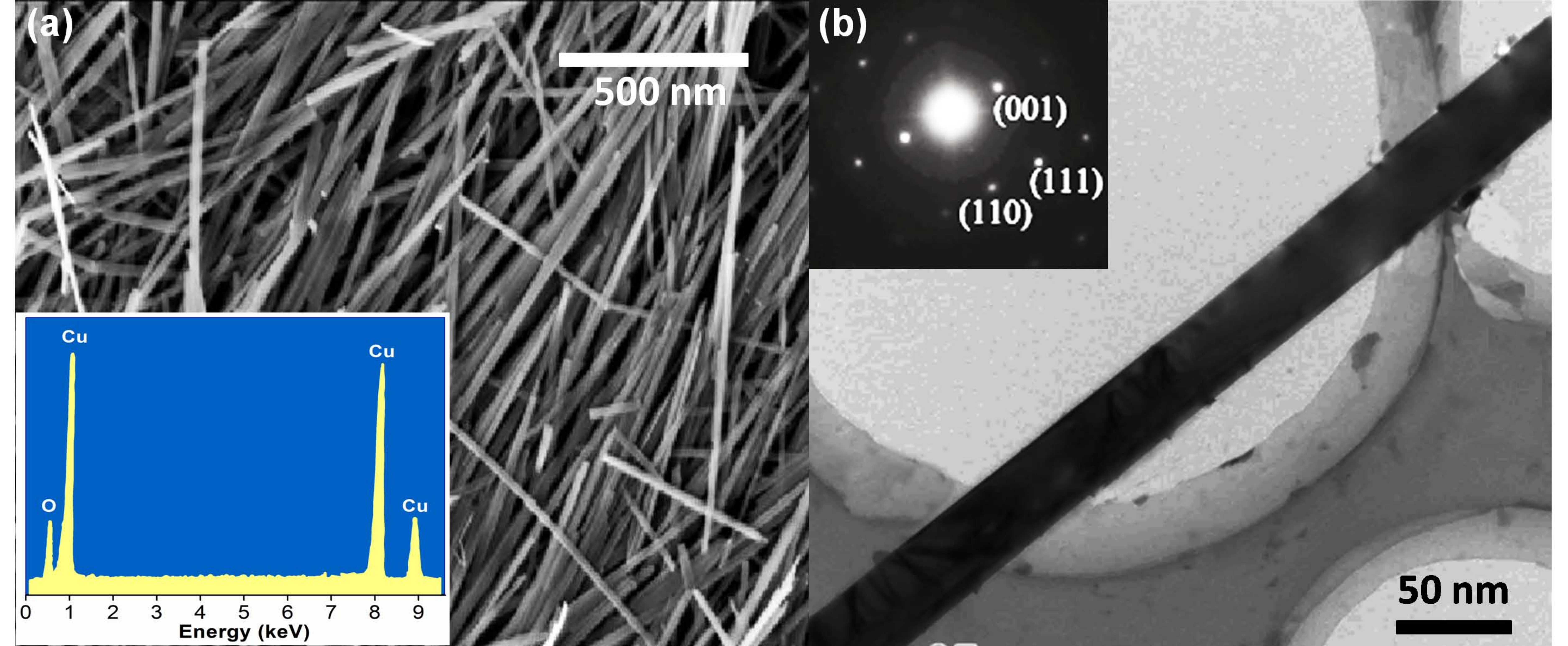}
\caption{\label{FIG.1} (a) SEM graph of grown Cu2O nanowires. The inset shows the energy dispersion x-ray analysis (EDX) results. (b) TEM picture of a single $\mathrm{Cu_2O}$ nanowire with the inset showing the corresponding selected area electron diffraction (SAED) pattern}
\end{figure}
\vfill
\newpage

\begin{figure}[H]
\centering
\includegraphics[width=0.8\textwidth]{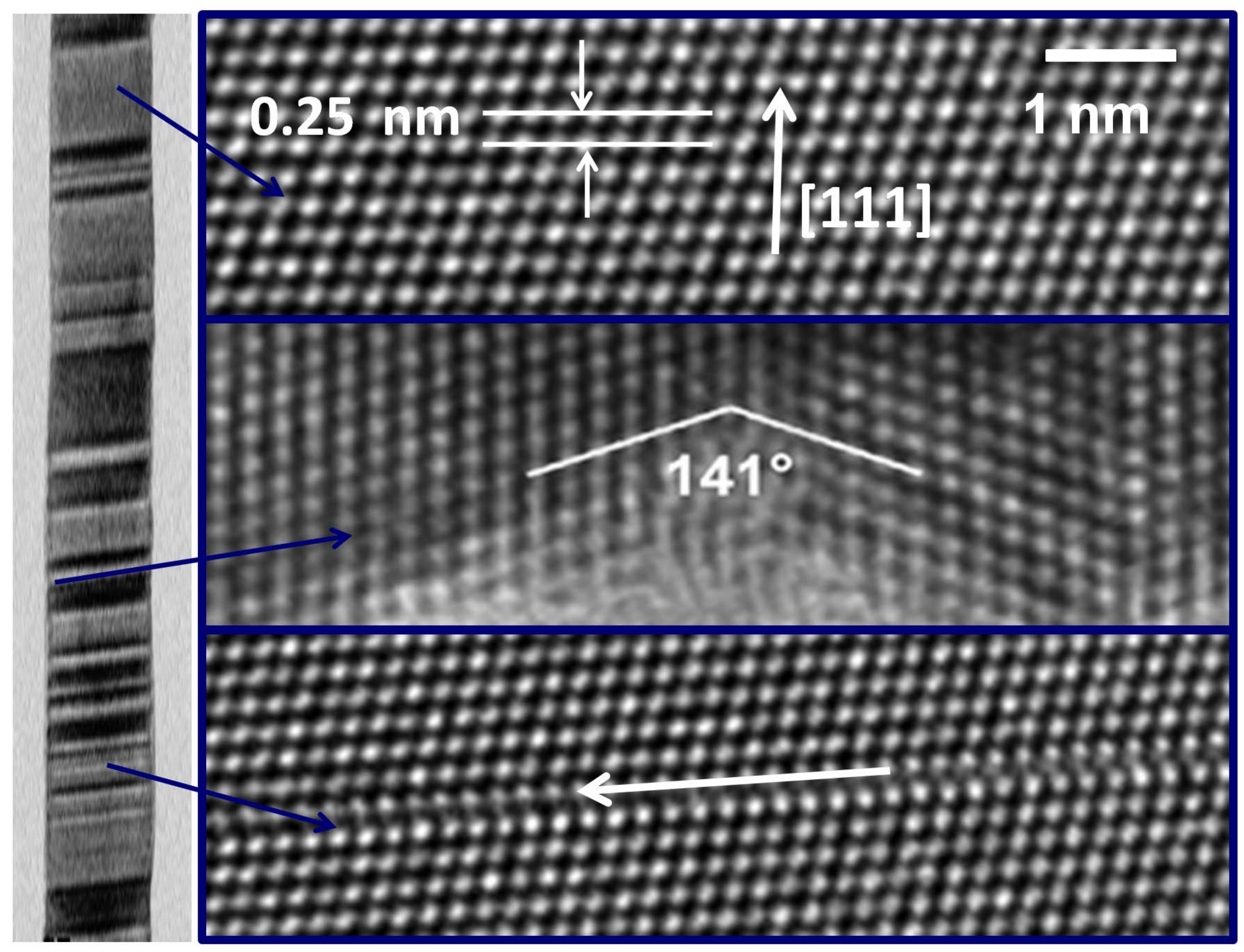}
\caption{\label{FIG.2} HRTEM images of three representative regions which contain no line-defect (upper), a twin structure (middle) and an intrinsic stacking fault (lower) respectively.}
\end{figure}
\vfill
\newpage
\begin{figure}[H]
\centering
\includegraphics[width=1\textwidth]{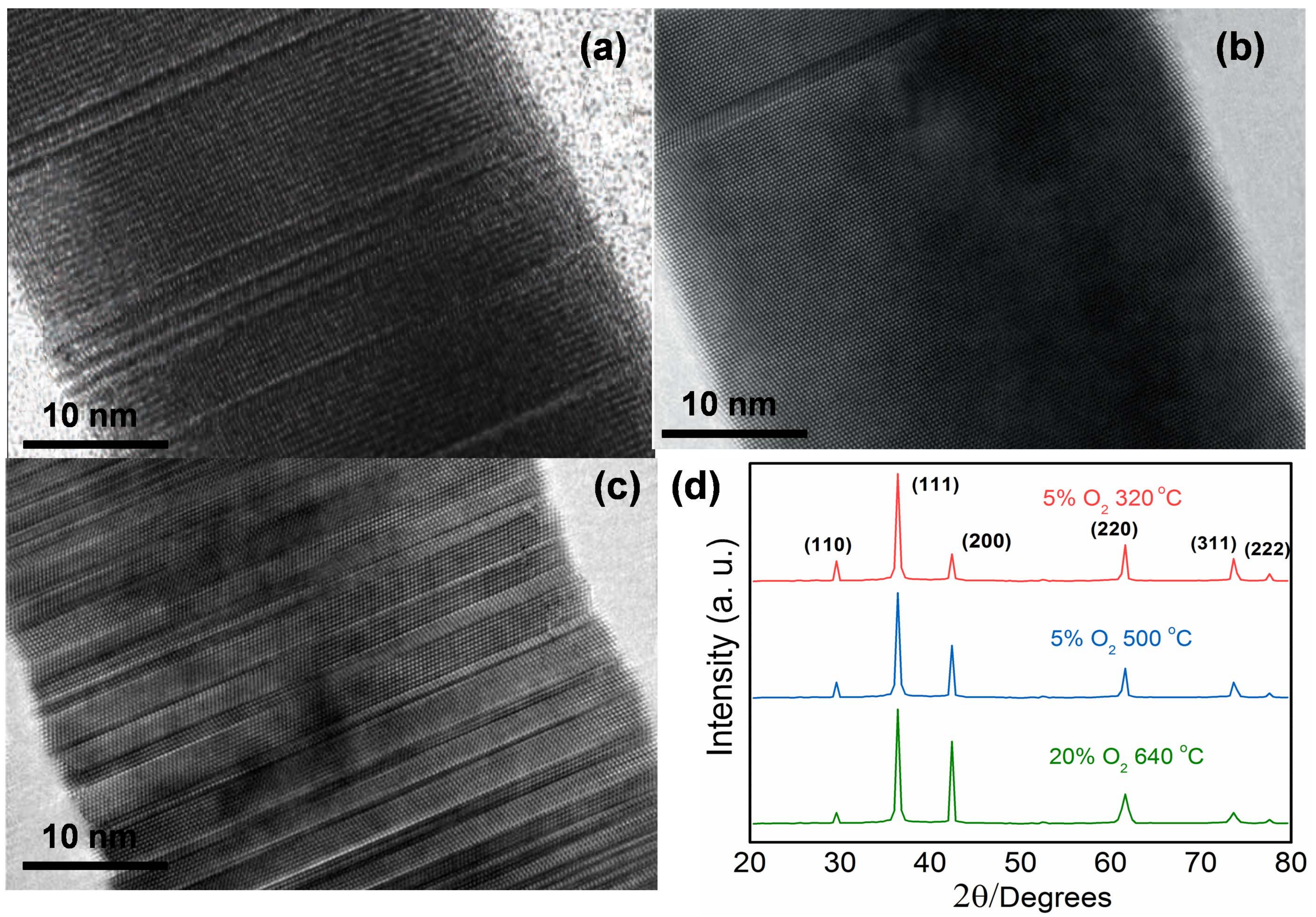}
\caption{\label{FIG.3} HRTEM images of nanowires grown at (a) 5\% $\mathrm{O_2}$ partial pressure and 320 \textcelsius, (b) 5\% $\mathrm{O_2}$ partial pressure and 500 \textcelsius{} and (c) 20\% $\mathrm{O_2}$ partial pressure and 640 \textcelsius, and all the NWs are oxidized for 5 hours; (d)  X-ray diffraction (XRD) curves for these three samples.}
\end{figure}
\vfill
\newpage

\begin{figure}[H]
\centering
\includegraphics[width=0.9\textwidth]{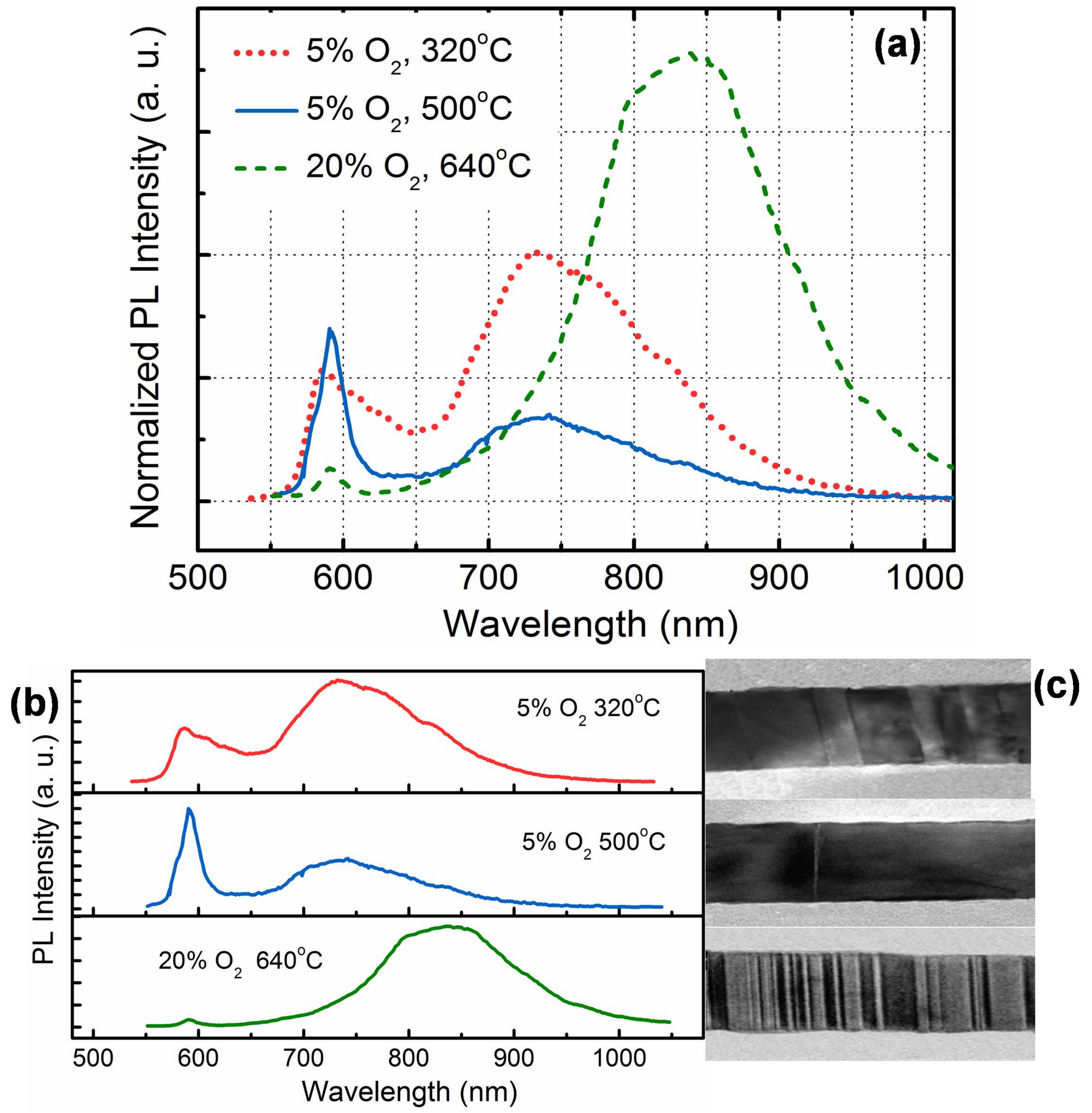}
\caption{\label{FIG.4} (a) Normalized low-temperature (77K) PL spectra for all three samples; (b) Individual PL spectrum for each sample with its representative TEM image shown in (c)}
\end{figure}

\bibliography{}

\end{document}